\documentclass[prl, twocolumn, floatfix]{revtex4}
\setcitestyle{super}

\usepackage{graphicx}
\usepackage{hyperref}
\usepackage{amsmath}
\usepackage{color}
\usepackage[normalem]{ulem}
\begin{document}

\title{Synchronization of a Self-Sustained Cold Atom Oscillator}

\author{H.\ Heimonen$^{1}$, L.~C.\ Kwek$^{1,2,3,4}$, R. Kaiser$^{5}$ and G.\ Labeyrie$^{5}$\footnote{To whom correspondence should be addressed.}}
\affiliation{$^{1}$Centre for Quantum Technologies, National University of Singapore, 3 Science Drive 2, Singapore 117543}
\affiliation{$^{2}$Institute of Advanced Studies, Nanyang Technological University, 60 Nanyang View, Singapore 639673}
\affiliation{$^{3}$National Institute of Education, Nanyang Technological University, 1 Nanyang Walk, Singapore 637616}
\affiliation{$^{4}$MajuLab, CNRS-UNS-NUS-NTU International Joint Research Unit, UMI 3654, Singapore}
\affiliation{$^{5}$Universit\'{e} C\^{o}te d'Azur, CNRS, Institut de Physique de Nice, 06560 Valbonne, France}

\begin{abstract}

Nonlinear oscillations and synchronisation phenomena are ubiquitous in nature. We study the synchronization of self oscillating magneto-optically trapped cold atoms to a weak external driving. The oscillations arise from a dynamical instability due the competition between the screened magneto-optical trapping force and the inter-atomic repulsion due to multiple scattering of light. A weak modulation of the trapping force allows the oscillations of the cloud to synchronize to the driving. The synchronization frequency range increases with the forcing amplitude. The corresponding Arnold tongue is experimentally measured and compared to theoretical predictions. Phase-locking between the oscillator and drive is also observed. 

\end{abstract}

\date{\today}
\maketitle

{\bf Introduction.---}Ever since the Dutch scientist, Christiaan Huygens, observed synchronization in two coupled pendulums hung from a common support, synchronization phenomena have been observed pervasively in numerous different settings, ranging from cardiac pacemaker to circadian rhythms to entrainment of pulsatile insulin secretion and locked states in laser systems \cite{Strogatz2001, Pikovsky2001}. Synchronization of geodesic acoustic modes and magnetic fluctuations in toroidal plasmas have also recently been observed for the first time\cite{Zhao2016}. It has been suggested that these observations provide a study of the nonlinear interactions among magnetic islands and the low frequency zonal flows. 

Synchronization has been an extremely useful and powerful tool for the analysis of a plethora of problems related to interactions between nonlinear oscillators \cite{Arenas2008} including various applications to phase locking and control in laser systems, chaos based communications \cite{Argyris2005}, and coupled chaotic oscillators \cite{Schroder2015}. Characterization of synchronization has also been made possible through a number of order parameters and information theoretic measures, including cross-correlation, a Green's function approach \cite{Schroder2015,Yokoshi2017}, degree of phase coherence \cite{Zhang2015} and mutual information. 

A central requirement for synchronization is the existence of self-sustained oscillations \cite{Pikovsky2001,Balanov2009}. Self-sustained oscillations occur ubiquitously in nature and are considered a subset of a larger class of dynamical systems. The main feature of self-sustained oscillators is that these systems continue to oscillate in their own rhythm after they are isolated or taken apart from a combined system. They are mathematically modelled using limit cycles\cite{Strogatz2001, Pikovsky2001, Balanov2009}. 

With the emergence of laser cooling techniques, collective behavior of ultracold atomic molasses in magneto-optical traps (MOT) has been observed and studied. These experiments have revealed dynamic instabilities associated with the appearance of self-sustained radial oscillations.  These oscillations have been found to be the result of the competition between a long range scattering interaction and the MOT's confining force \cite{unstableMOT}, providing an excellent platform for the study of plasma and astrophysical phenomena such as pulsating stars.

In this paper we present a simple theoretical model for the MOT used to explain the self-sustained oscillations and synchronization to an external drive. We then describe the experimental procedure to observe the oscillations and synchronization. Finally we analyse the data and compare the Arnold tongue \cite{Pikovsky2001} obtained from the theory to the one obtained from experiment and comment on the differences.

We use a simplified model of a magneto-optical trap, considering the trap to be spherically symmetric and the gas to be of constant density throughout the trap\cite{unstableMOT}. More detailed models exist\cite{}, but the simple model suffices for our purposes. In this model, an atom at the edge of the cloud will experience a force due to the laser directly facing the atom and an attenuated force due to the laser from the opposite side of the cloud. In the low intensity Doppler model, the magneto optical force, with saturation parameter $s_{inc}$, on an atom at location $x$ and velocity $v$ on or outside the edge of a cloud of radius $R$ is \cite{unstableMOT, metcalf2012}:

\begin{align}
\begin{split}
F_{trap}(x,v) = {}& -\frac{\hbar k \Gamma}{2} s_{inc} \frac{1}{1+\frac{4(\delta_0+\mu\nabla B x+kv)^2}{\Gamma^2}} \\
{}&+\frac{\hbar k \Gamma}{2} s_{inc} \frac{e^{-b}}{1+\frac{4(\delta_0-\mu\nabla B x-kv)^2}{\Gamma^2}} \\
{}&+\eta \frac{\hbar k \Gamma}{2} s_{inc} \frac{1}{1+\frac{4 \delta_0^2}{\Gamma^2}}(1-e^{-b})\bigg(\frac{R}{x}\bigg)^2
\label{force}
\end{split}
\end{align}

The first term of the equation corresponds to the force of the laser detuned by $\delta_0$ from the atomic transition directly incident on atoms with Zeeman shift $\mu\nabla B x$ (for magnetic field gradient $\nabla B$), Doppler shift $kv$, and atomic transition linewidth $\Gamma$. This beam pushes the atoms inward towards the centre of the cloud. The second terms corresponds to the laser beam incident from the opposite side of the cloud, attenuated by the amount $e^{-b}$ due to scattering in the cloud of optical thickness $b$. The third term in the equation is the repulsive force between atoms due to the scattered photons\cite{Wieman}. By Gauss' law, this force outside the cloud must be proportional to $1/x^2$. $\eta$ is the fraction of the absorption cross-section of the incident laser frequency and that of the inelastically scattered light.

At the edge of the cloud ($x=R$) the forces balance to zero, but the model predicts that the edge undergoes a bifurcation above a critical radius $R_{crit}$. The instability threshold \cite{unstableMOT} is that the critical MOT detuning is equal to the Zeeman shift experienced by atoms at the edge of the cloud: $\left| \delta_{crit} \right| = \mu~\nabla B~R_{crit}$. The bifurcation creates a limit cycle in the phase-space portrait of the system, corresponding to spectacular "breathing" mode oscillations of the cloud expanding and contracting in volume. The limit cycle in the phase-space portrait of an autonomous equation of motion is a sufficient condition for the oscillator to be able to synchronize \cite{Pikovsky2001}. By adding a time-periodic weak driving term to the MOT we can establish control over the phase and frequency of the oscillations. We accomplish this by sinusoidally varying the detuning of the lasers according to $\delta(t) = \delta_0 + \Delta \cos(2 \pi \nu_f t)$ for $\Delta/\delta_0 \ll1$. With a weak driving strength we can series expand the force up to first order in $\Delta$ such that $F_{driven} \approx F_{trap}+\alpha\Delta\cos(2 \pi \nu_f t)$ for some proportionality constant $\alpha$. Thus modulating the detuning of the MOT acts like sinusoidally "squeezing" and "stretching" the trap with an external force.

\begin{figure}
\begin{center}
\includegraphics[width=1\columnwidth]{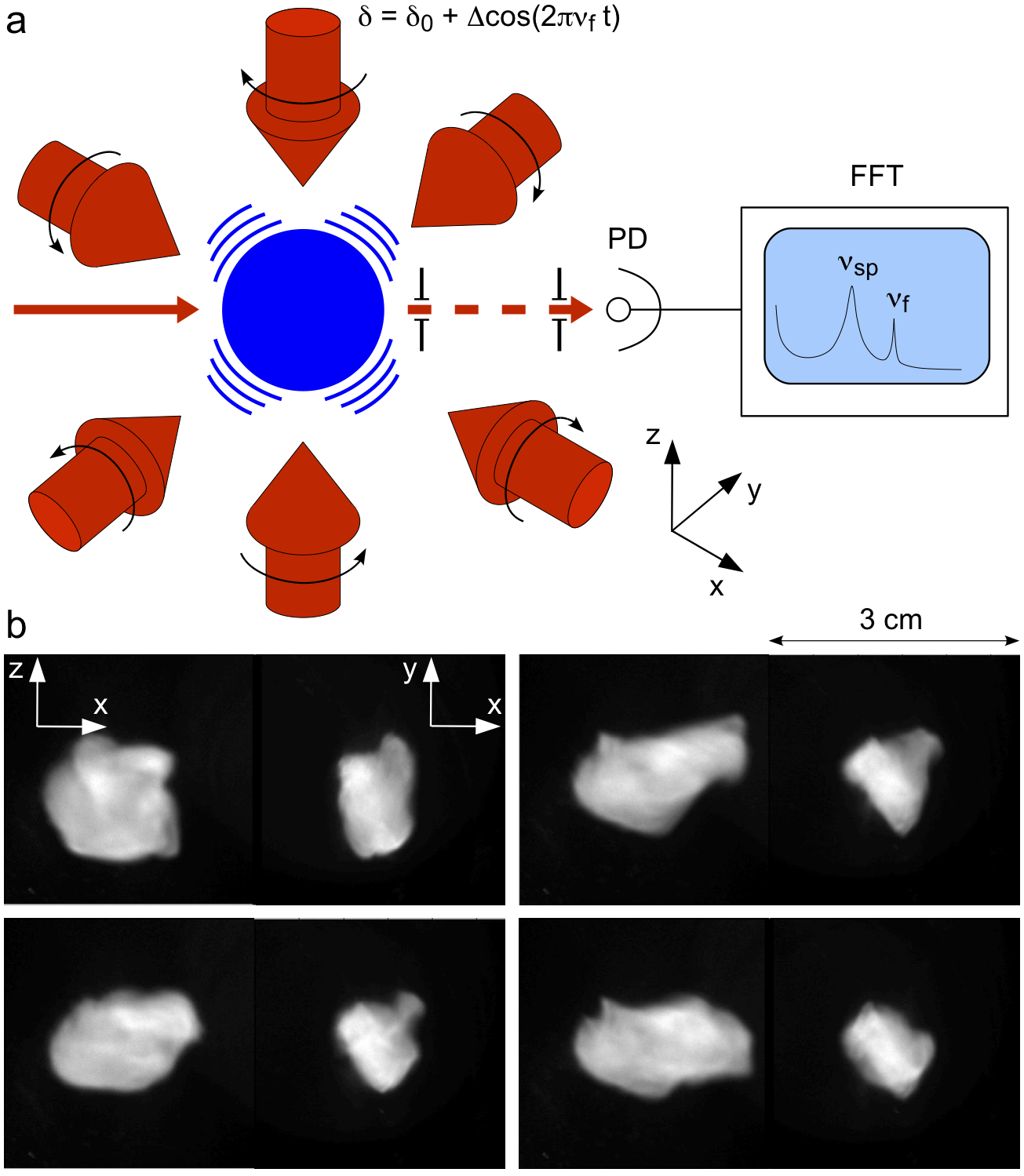}
\caption{a) Experimental setup with six independent laser beams and a small probe beam. b) Spontaneous oscillation mode of the MOT. Snapshots of the MOT fluorescence distributions along two lines of sight.}
\label{fig1}
\end{center}
\end{figure}

The principle of the experiment is sketched in Fig.~\ref{fig1} a. It is based on a large magneto-optical trap of $^{87}$Rb. The features of this device are described in detail in ref.~\cite{VLMOT}, we just outline its most important characteristics here. Satisfying $R > R_{crit}$ in the experiment requires a large number of trapped atoms. This vapor-loaded MOT has a large capture range due to the large size of the trapping beams ($1/e^2$ radius = 2.6 cm), which enable us to trap as much as $10^{11}$ atoms. Due to this large number, the interatomic repulsive force due to scattered photon reabsorption~\cite{Wieman} is strong and the corresponding size of the trapped cloud is large, typically $> 1$ cm. As the stable MOT radius $R$ typically scales as $1 / \nabla B$, it can be seen that the critical detuning is nearly independent of the magnetic field gradient\cite{unstableMOT}. For the parameters used in the present paper, $\delta_{crit} \approx -2 \Gamma$. Thus, assuming the cloud size is independent of detuning, (which is relatively reasonable in our situation\cite{VLMOT}), one sees that the MOT is unstable for detunings smaller (in absolute value) than the critical value, and stable for larger values. In the experiment, we approach the ideal situation of the lasers symmetrically arranged by using six intensity-balanced ''independent'' laser beams. In the different case of retro-reflected beams, the intensity imbalance between counter-propagating beams generated by the opacity of the cloud provokes other spatial instabilities that involve mainly the center-of-mass motion of the cloud\cite{Lille1, Lille2}. Note however that in real life, even in the independent beam configuration the instability is usually complex with both radial and center-of-mass motion, as seen in the snapshots of Fig.~\ref{fig1}b.

\begin{figure}
\begin{center}
\includegraphics[width=1\columnwidth]{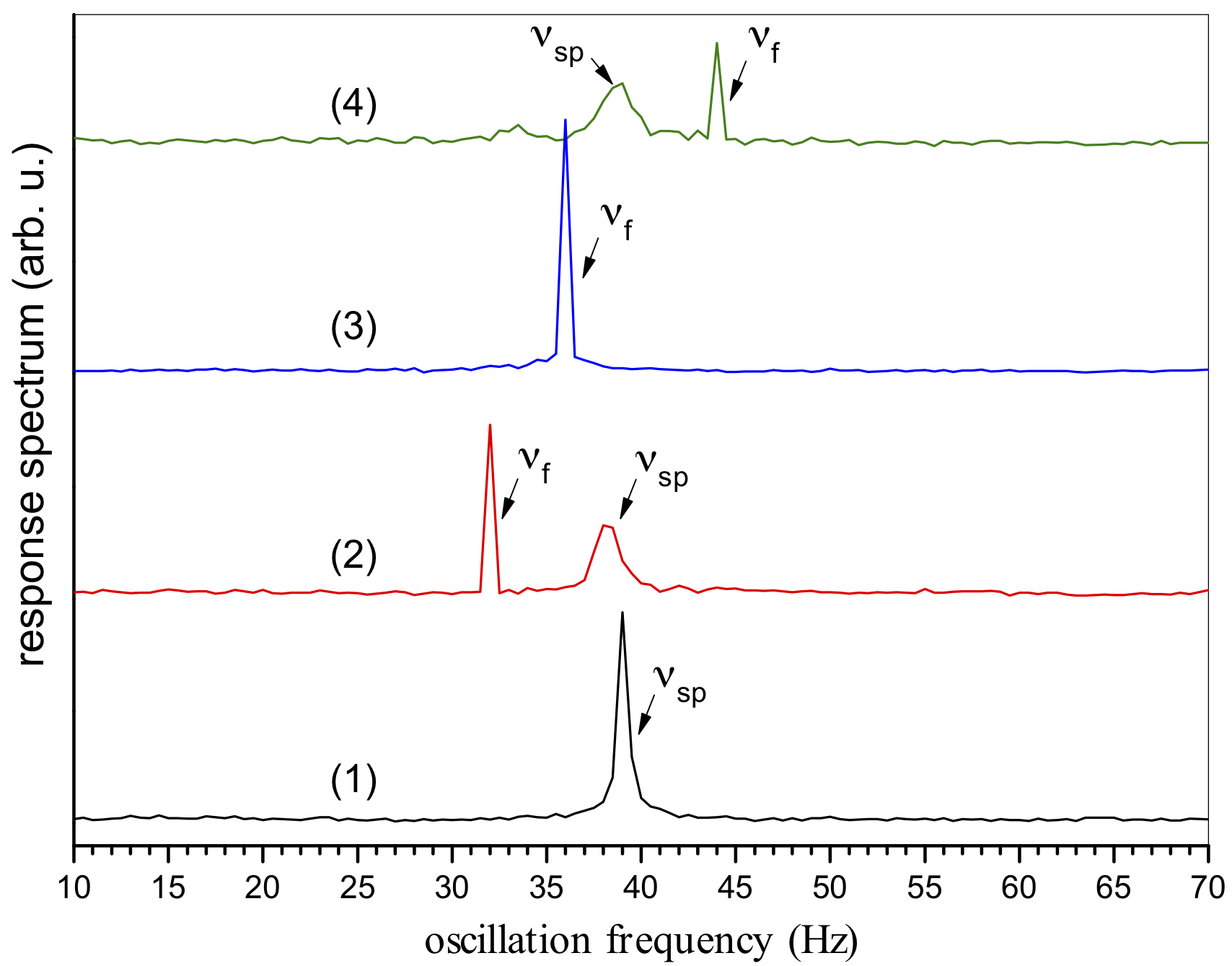}
\caption{Experimental oscillation spectra. (1) Spontaneous oscillation at $\nu_{sp} = 39$ Hz. (2) Unsynchronized forced oscillation with $\nu_f = 32$ Hz. (3) Synchronized forced oscillation with $\nu_f = 36$ Hz. (4) Unsynchronized forced oscillation with $\nu_f = 44$ Hz. $\Delta / \delta_0 = 0.067$.}
\label{fig2}
\end{center}
\end{figure}
 
We start the experiment with a self-oscillating MOT (detuning $\delta_0 = -2 \Gamma$, $\nabla B = 5$ G/cm, and $\mu = 1.4$ MHz/G), and apply the weak forcing by sinusoidally modulating the trapping laser detuning. We monitor the response of the cloud by measuring in real time the transmission of a weak probe laser beam, centred on the MOT, whose transverse size is smaller than that of the cloud (see Fig.\ref{fig1} (a)). This transmission is a measure of the number $N$ of atoms contained in the volume defined by the intersection of the probe beam and the cloud's spatial distribution, which varies with time as the cloud oscillates. Compared to a MOT fluorescence measurement, which is intrinsically sensitive to a modulation of $\delta$, this detection scheme has the advantage to be sensitive to $N(t)$ only. 

To analyse the data for frequency entrainment, we perform a Fourier transform of the transmission signal to obtain spectra such as those shown in Fig.~\ref{fig2}. Spectrum (1) corresponds to spontaneous oscillations, without external modulation ($\Delta = 0$). The motion is observed to be quite monochromatic, with a central frequency $\nu_{sp} = 39$ Hz. We stress that such a monochromatic mode is only observed in a narrow parameter range just above the instability threshold: decreasing $\left| \delta \right|$ results in more chaotic oscillations, with a broad low frequency pedestal in the spectra. Spectrum (3) corresponds to an external modulation at 36 Hz, with $\Delta / \delta_0 = 0.067$. The spontaneous peak has disappeared and the MOT now oscillates at the driving frequency: this is considered synchronized motion. When the driving frequency is detuned by a large amount from the spontaneous frequency (spectra (2) and (4)), the spontaneous peak reappears and the system oscillates at both frequencies, corresponding to the unsychronized regime. Thus, the criterion for synchronization we will use in the following is the presence or absence of the spontaneous peak in the spectrum. This is chosen, because it provides a clear distinction between synchronized and unsynchronized signals. The criterion is stricter than necessary from the theoretical point of view, as a synchronized oscillator in the presence of noise\cite{Stratonovich2} in general will have a spectrum like Fig.\ref{fig2}, (2) and (4). Nevertheless, the criterion is clear and sufficient, so we have adopted it for the purposes of the experiment.

\begin{figure}
\begin{center}
\includegraphics[width=1.0\columnwidth]{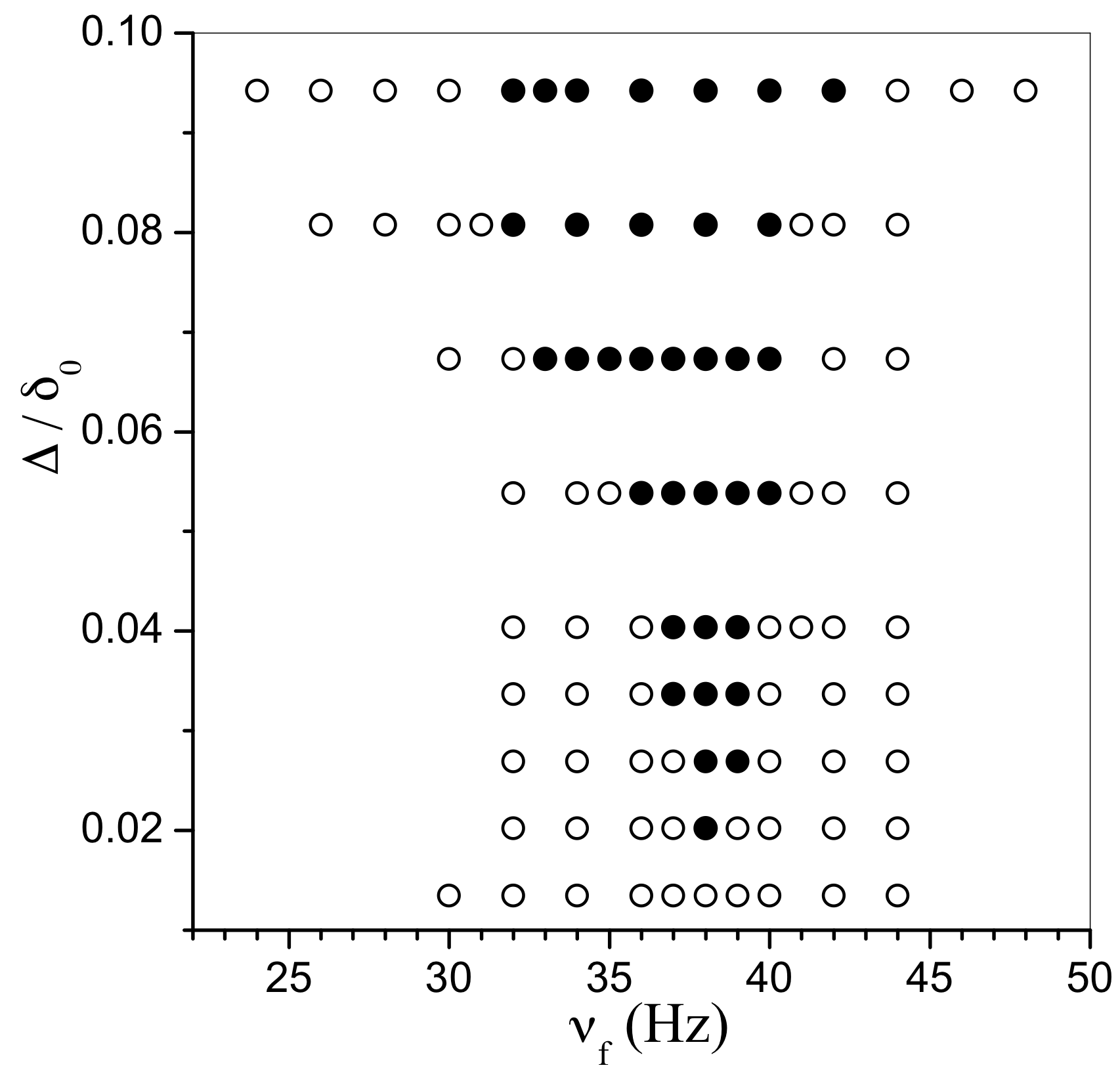}
\caption{Experimental Arnold tongue. We measure the synchronization range against forcing amplitude. Filled and open symbols correspond to synchronized and unsynchronized oscillations respectively.}
\label{fig3}
\end{center}
\end{figure}

\begin{figure}
\begin{center}
\includegraphics[width=1.0\columnwidth]{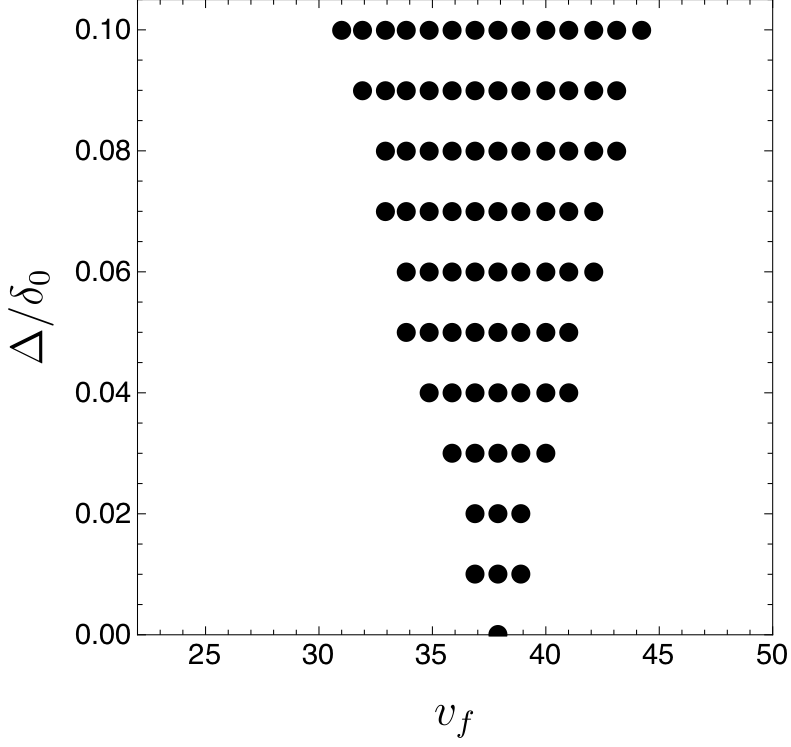}
\caption{Arnold tongue calculated from equation \ref{force} for the oscillator. The tongue corresponds to a noiseless oscillator, so it touches the $\nu_f$-axis and has clearly defined edges. }
\label{th_tongue}
\end{center}
\end{figure}

In Fig.\ref{fig3}, we report an experimental measurement of the Arnold tongue corresponding to our oscillator. The Arnold tongue\cite{Pikovsky2001, Balanov2009} represents the region of driving frequencies $\nu_f$ for which the oscillator synchronises to the frequency of the drive, plotted over a range of weak driving strengths $\Delta / \delta_0$. The Arnold tongue computed from Eq. \ref{force} is shown in Fig.\ref{th_tongue}. The most notable differences between the tongues are that the theoretical model predicts the sync region to touch the $\nu_f$-axis and be symmetric about the natural frequency. In addition it is clear that for very weak driving the width of the Arnold tongue is narrower in the experiment than the theoretical calculation. The differences between the experiment and theory are consistent with the presence of noise in the system \cite{Pikovsky2001}, as noise will effectively mask the presence of very weak driving and cause phase slips when close to the edge of the region. As a result, a very weak driving force will be masked by the noise and no synchronization is observed. The difference in the shape of the experimental and theoretical Arnold tongues must be attributed to the simplified theoretical model used for the calculations. Nevertheless the experiment agrees well with theory when considering real-world limitations.

\begin{figure}
\begin{center}
\includegraphics[width=1.0\columnwidth]{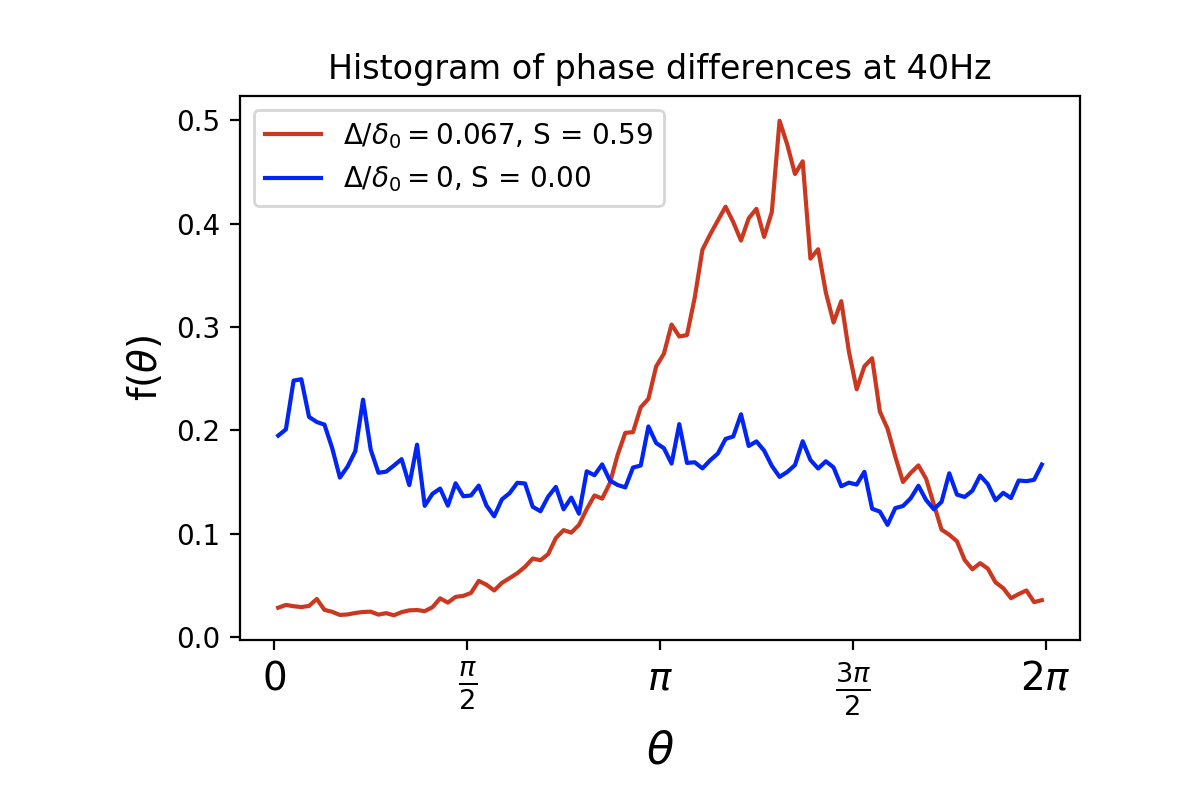}
\caption{Histogram of phase difference of the external driving at 40 HZ and the oscillator at zero driving strength and at a fixed strength $\Delta / \delta_0 = 0.067$. The natural frequency of the oscillator $\nu_{sp} = 39$Hz. From the histogram we can calculate the syncronization measure S for the two cases to show the presence of synchronization in the system.}
\label{hist}
\end{center}
\end{figure}

The Arnold tongue is a tool to investigate the frequency entrainment aspect of synchronization, but we can also study the phase aspects of the oscillator. Instead of Fourier-transforming the signal to study the frequency components, we perform a Hilbert-transform \cite{Pikovsky2001}, to extract the instantaneous phase of the oscillator at all times. We can then compute the phase difference $\theta = \phi_{osc}-\phi_{drive} \mod 2\pi$ and plot a histogram of all the observed phase differences during a run of the experiment. The histogram is shown in Fig. \ref{hist} for a driving at 40Hz, when the natural frequency of oscillation is 39Hz. The width of this peak can be characterised using the first trigonometric moment\cite{mardia} S, which is a well-behaved measure of phase synchronization \cite{Kurths}. The value $S=0$ corresponds to a perfectly flat phase distribution $f(\theta)$, whereas the maximal $S=1$ corresponds to a delta-function. We note, that it is necessary to use directional statistics for the case of phase, which is a $2\pi$ periodic property. 
\begin{equation}
S = \Big|\frac{1}{2\pi}\int_{0}^{2\pi}f(\theta) \ e^{i\theta}d\theta\Big|
\label{s}
\end{equation}
When the drive is not coupled to the oscillator ($\Delta/\delta_0 = 0$), the phase of the oscillator is completely uncorrelated with the phase of the drive, presenting a flat distribution corresponding to equal likelihood for all phase values. This distribution has a synchronization measure of $S=0$. When the driving strength is increased to $\Delta/\delta_0 = 0.067$ the phase of the oscillator becomes correlated with the drive and the phase distribution becomes peaked. For this distribution $S= 0.59$, indicating a relatively high degree of synchronization in the presence of noise.

In conclusion, we have experimentally shown synchronization of a self-sustained cold-atom oscillator to an external drive, and compared the results to theory. The data show frequency entrainment of the oscillator to the drive, and phase coherence between the oscillator and the drive when in the synchronization region. The results are further evidence that the oscillations are a fundamental property of the MOT, due to limit cycle oscillations of large cold atom clouds. Synchronization of the cloud to an external modulation provides new insight in the study of MOT instabilities, and establishes control over the frequency and phase of the oscillations. This control can be used in further studies of MOT instabilities, for example to increase the density of the atomic cloud by turning off the MOT when the cloud is at its minimum size. Timing the turn off can be done by phase locking the oscillator to a drive first and using the phase of the driving signal. The method might also be extended to stabilising unstable traps by applying the driving signal $\pi$ out of phase with the oscillations, though this would require stronger driving than was considered in this work. The technique can also be of interest in the study of oscillating plasmas due to the close analogue between the systems.

{\large \bf Acknowledgements}\\
{\footnotesize We acknowledge stimulating discussion with D. Wilkowski, A. Chia and M. Hajdusek. The Sophia Antipolis group is supported by CNRS, UNS, and R\'{e}gion PACA. The CQT group is  supported  by  the  MOE  grant  number RG 127/14, and the National Research Foundation, Prime Minister's Office, Singapore under its Competitive Research  Programme  (CRP  Award  No.   NRF-CRP-14-2014-02)

\noindent

\end{document}